\documentclass[11pt]{article}
\usepackage{amsmath}
\usepackage{graphicx}
\usepackage{latexsym}
\usepackage{amssymb}

\numberwithin{equation}{section}

\begin{document}

\title{ Charge without charge in  quarks}
\author{ Harry Schiff \\Professor Emeritus of
              Physics,
University of Alberta\\
Edmonton, AB, Canada, T6G 2J1\thanks{email: hschiff@shaw.ca}}

\date{}
\maketitle

 \noindent {\small PACS numbers:  12.39.-x}

\begin{abstract} With appropriate gauge transformations, field can replace electric charge in quarks. Classical quarks, in a necessary non-gauge invariant formulation, are used for illustration, bringing to the fore the limitations of the usual electric charge densities for single particles in Coulomb equations. The results are encouraging; the solutions for the Coulomb potentials apply individually to each quark in a shell structure. A remarkably simple relation emerges between the Coulomb and weak potentials.
\end{abstract}

\section{Introduction}

	To introduce the necessary equations  and realize the idea that field can replace electric charge in quarks, this article extracts and adapts a portion of a previous article on classical quarks [1]. 

	The first part examines the homogeneous equations with solutions for strong and weak potentials, highlighting the importance of the latter for describing the arrangement of quarks in hadrons; the second part focuses on the particular or Coulomb solutions, where the non-gauge invariance of the system allows for non-trivial gauge transformations to confirm that `charge is a manifestation of field'.

Starting with the Li\'{e}nard-Wiechert exact solutions of Maxwell's equations of a point charge $q$ in arbitrary motion, it can be shown that the following relation between the fields and potentials holds everywhere:  $[e = c= 1]$			
\begin{equation}	
(B^2 - E^2)  =  -q^{-2}(A^2 - \phi ^2)^2	
\end{equation}		
With $F_{\mu \nu}  = \partial _\mu A_\nu - \partial _\nu A_\mu $, 3-potential vector ${\bf A}, A_4 = i\phi$  and the well-known relation $F_{\mu \nu}  F_{\mu \nu}  = 2(B^2 - E^2)$, (1.1) can be written
\begin{equation}
F_{\mu \nu}  F_{\mu \nu}  =  -2q^{-2}(A_\mu A_\mu )^2				
\end{equation} 
The Li\'{e}nard-Wiechert solutions also satisfy the orthogonal relation $E\cdot B = 0$ everywhere, i.e.,
\begin{equation}{}^\ast\!F_{\mu \nu}  F_{\mu \nu}  = 0	
\end{equation}		
In (1.3) the familiar notation ${}^\ast\!F_{\mu \nu}  =   1/2i\varepsilon_{\mu \nu \alpha \beta}F_{\alpha \beta}$  is used for the dual of $ F_{\mu \nu}  $  using the Levi-Civita tensor $\varepsilon_{\mu \nu \alpha \beta}$.	
	
Due to the opposite symmetries of $ F_{\mu \nu}  $ and $A_{\mu}A_\nu$, (1.2) can be written,	
\begin{equation}  (F_{\mu \nu}+ \surd 2q^{-1}A_\mu A_\nu )^2  = 0	
\end{equation}			 
In the following the factor $\surd 2$ will be absorbed by the potentials.				Defining the mixed tensor
\begin{equation}
W_{\mu \nu}  \equiv F_{\mu \nu}  + q^{-1}A_\mu A_\nu,	
\end{equation}			 
 (1.4) can also be written 
\begin{equation}
W_{\mu \nu}  W_{\mu \nu}  = 0
\end{equation}  
Further, noting that ${}^\ast W_{\mu \nu}  =  {}^\ast\!F_{\mu \nu}  $, since $ \varepsilon_{\mu \nu \alpha \beta}A_\alpha A_\beta   =  0$, (1.3) can be written,				 
\begin{equation}
{}^\ast W_{\mu \nu} W_{\mu \nu}= 2E\cdot B = 0 	
\end{equation}
The expressions (1.6) and (1.7) show that $W_{\mu \nu}$ (1.5) satisfies formally the conditions for a null field (although the term `null field' applies strictly to an anti-symmetric tensor, we have taken the liberty to use it here for the mixed tensor)	 that holds everywhere in space and time. This interesting property of $W_{\mu \nu}$, a combination of the exact solutions for the electric and magnetic fields and potentials of Maxwell's equation for a point charge, arouses curiosity as to whether $W_{\mu \nu}$ may exhibit other features of interest.  To this end we consider the non-gauge invariant equation
$\partial _\mu W_{\mu \nu}=  -4\pi{}^e\!j_\nu$, 
\begin{equation} \partial _\mu( F_{\mu \nu} + q^{-1}A_\mu A_\nu)  =  -4\pi{}^e\!j_\nu
\end{equation}
where ${}^e\!j_\nu$ is an external localized 4-current density $ ({}^e\!{\bf J} , i {}^e\!\rho)$  that at present can be chosen freely. 

\section{Solutions of (1.8)}
 
Only static, radially symmetric solutions of (1.8) are considered,
 \begin{equation}       
q^{-1}\nabla \cdot (A{\bf A})  =  -4\pi \, {}^e\!{J}(r) 
\end{equation}			
\begin{equation}	
\nabla\cdot {\bf E} - q^{-1}\nabla \cdot (\phi {\bf A})  = 4\pi \, {}^e\!\rho(r)			
\end{equation}
For the asymptotic solution of (2.1),
\begin{equation}
\nabla\cdot (A{\bf A})  = 0,				
\end{equation}
the 3-potential,			
\begin{equation}	
A = a/  r 	  				
\end{equation}
with arbitrary charge $a$.

Defining the dimensionless parameter	
\begin{equation}\gamma \equiv q^{-1}a,
\end{equation}
\begin{equation}A = \gamma q/r
\end{equation}
Using (2.6), asymptotically (2.2) can be written,		
 \begin{equation}
\frac{d^2 (r \phi)}{dr^2} + \frac{\gamma d (r \phi)}{rdr}=0
\end{equation}
For any value of $\gamma$, except $\gamma = 1$, treated separately below, there are two indicial solutions of (2.7),
\begin{equation}\phi _1 = b/ r	
\end{equation}
\begin{equation}\phi_2 = cr^{ -\gamma}	
\end{equation}
Showing a Coulomb potential with charge $b$ and $a$ possible absolute confining potential for 
$\gamma < 0$; both $b$ and $c$ may be positive or negative. 	For $\gamma =1$ the two indicial solutions of (2.7) merge to a single Coulomb potential so a second solution is needed. This is easily seen to be given by $r\phi  \sim   \log r$. This solution will not be used here.

\section{Null Condition and Fractional Charges} 
 As both $A$ and $\phi$ go as $1/r$ asymptotically, continuing with an application of (1.2) to these solutions, one obtains an equation for the charge $b$. Thus, with $B = 0$, (1.2) becomes
\begin{equation} - E^2 + q^{-2} (A^2 - \phi^2)^2 = 0	
\end{equation}
For each sign of $q$ one obtains a quadratic equation for the Coulomb charge $b$ in (2.8),
\begin{equation}b^2 \pm qb - \gamma^2q^2/2  = 0
\end{equation}
Thus for the quadratic with the minus sign,
\begin{equation}    b^2 - qb - \gamma^2q^2/2  =  0,	
\end{equation}
The charge $b$ has two solutions, 
\begin{equation}b (\pm)  = q[1 \pm \surd (1+2\gamma^2)]/2	
\end{equation}	
Choosing $\gamma^2 = 4$ in (3.4) and $q = \pm1/3 $, one gets 
\begin{equation}b (\pm) = \pm(2/3, -1/3),   
\end{equation}	
Confinement follows in (2.9) with $\gamma = -2$. 

\section{Lagrangian}
		
A Lagrangian for (1.8) cannot be obtained using only the field components in that equation since the nonlinear addition to the second order Maxwell terms of the potentials in (1.8) is in first order, similar to a dissipative term. This is resolved with the addition of an 
electromagnetic field\footnote{See, for example, Morse \& Feshbach, {\em Methods of Theoretical Physics} (McGraw-Hill Book Company, Inc.,  1953), Vol.\ 1, p 298, 313. Where they discuss introducing mirror image systems for dissipative systems.} $G_{\mu \nu}$ and its associated potential $V_\mu$,
\begin{equation} G_{\mu \nu}= \partial _\mu V_\nu - \partial _\nu V_\mu	
\end{equation}
The following Lagrangian includes $ F_{\mu \nu}, A_\nu$ and $ G_{\mu \nu}, V_\nu$ and their respective currents ${}^e\!j_\nu$ and ${}^\ast\!j_\nu $: 
\begin{equation}                 L  =  -1/8\pi [1/2 G_{\mu \nu} F_{\mu \nu}+ q^{-1} V_\nu\partial_\mu (A_\mu A_\nu) + 4\pi V_\nu {}^e\!j_\nu + 4\pi A_\nu {}^\ast\!j_\nu]      
\end{equation}
Variation of $L$ with respect to $V_\mu $ gives (1.8) while variation with respect to $A_\mu $ yields a linear equation for $V_\mu $.  
\begin{equation}              \partial _\mu G_{\mu \nu}- q^{-1} A_\mu  (\partial _\mu V_\nu + \partial _\nu V_\mu)  =  -4\pi^\ast\!j_\nu	
\end{equation}		
From (4.3) the equation for $V_4 = i\psi$ becomes,
\begin{equation}\frac{d^2\psi}{dr^2}  +  \frac{2d\psi}{rdr}  -  q^{-1} \frac{Ad\psi}{dr}   =  -4\pi ^\ast\!\rho(r) 		 
\end{equation}
For the Coulomb energy, considered below, the particular solution of (4.4), $\psi_p$, will be chosen to be equal to the particular solution $\phi_p$ of (2.2), i.e. $\psi_p  =  \phi_p$.

From the Lagrangian (4.2) an expression for the canonical stress-energy tensor,   
\begin{equation}T_{\alpha \beta}   =   - \frac{\partial L}{\partial (\partial _\alpha A_\lambda)
} \partial _\beta A _\lambda -
 \frac{\partial L }{\partial (\partial _\alpha V_\lambda ) }  \partial_\beta V_\lambda   + L\delta_{\alpha \beta} 
\end{equation}
Although the Lagrangian functions for obtaining the equations, $T_{\alpha \beta}$ is not satisfactory for a proper canonical stress-energy tensor. The external sources are unknown and the diverging strong potential precludes integration for a total energy. Instead, the energy of a quark will be taken as consisting of two distinct parts, the Coulomb potential energy and separately, the contribution from the strong and weak potentials. For the former we choose from (4.2) the stress-tensor of the electromagnetic field involving $F_{\alpha \mu}$ and $G_{\alpha \mu}$:
\begin{equation}\Theta_{\alpha \beta}  =  1/8\pi [G_{\alpha \mu}F_{\beta \mu} + F_{\alpha \mu}G_{\beta \mu} - 1/2(G_{\mu \nu}F_{\mu \nu})\delta_{\alpha \beta} ]
\end{equation}
With $\psi_p  =  \phi_p, F_{\alpha \mu} = G_{\alpha \mu}$ and the stress-energy density becomes,
\begin{equation}      -\Theta_{44}  = 1/8\pi [-2F_{4\mu}F_{4\mu} + 1/2 F_{\mu \nu}F_{\mu \nu}] =  1/8\pi(d\phi_p/dr)^2		
\end{equation}
The integrated Coulomb energy,
\begin{equation} \xi _c = 1/2\int (d\phi_p/dr)^2r^2dr	
\end{equation}
The strong and weak potentials, solutions of the homogeneous equations from (2.2) and (4.4), are designated as $\phi_k$ and $\psi_\lambda   $ where $k$ and $\lambda$ are the strong and weak potential charges respectively.

The quark energy from these potentials is chosen as 
\begin{equation}\xi_p  = b(\phi_k + \psi_\lambda),
\end{equation}	
with a quark fractional charge $b$.

  In the following the term `dimensionless' applied to a potential will mean the potential minus its dimensional constants. It is convenient to define the dimensionless vector potential $\mathcal{A} = q^{-1}A$, so that from (2.2) one can write the linear homogeneous equation for the 
strong potential $\phi_k$,
\begin{equation}{\rm Div}(d \phi_k/dr + \phi_k\mathcal{A}) = 0
\end{equation}
Thus in general, one can write, 
\begin{equation}
 d\phi_k /dr + \phi_k \mathcal{A} = Kk/r^2, 	
\end{equation}
with dimensionless parameter $K$.
 From (4.4) one has the homogeneous equation for the weak potential,
\begin{equation}\frac{d^2 \psi_\lambda }{dr^2}+\frac{2d \psi_\lambda }{rdr}
-\frac{Ad \psi_\lambda }{dr}=0
\end{equation}
Asymptotically, from (4.11) with $\mathcal{A} = \gamma/r$ (2.6) and $\gamma = -2$,
\begin{equation}d \phi_k/dr - 2 \phi_k/r  = 0
\end{equation}
Thus, for all $\phi_k  $ solutions, $\phi_k  \sim kr^2/s^3 +$ const, where $s$ is a length dimension.  Similarly, asymptotically (4.12) becomes,	   
\begin{equation}
\frac{d^2\psi_\lambda}{dr^2}  +  \frac{4d\psi_\lambda}{rdr}   =  0   
\end{equation}
For all $\psi_\lambda$ solutions,  $\psi_\lambda \sim \lambda s^2/r^3 +$ const.

\section{Multi-Quark Solutions}

	The homogeneous solutions for $\phi_k$ from (2.2) would require $A$ from (2.1).  But (2.1) is incomplete, lacking a magnetic field  from the use of spherical symmetry. Instead, $A$ (now $\mathcal{A}$) will be taken from (4.12),
\begin{equation}
\mathcal{A}= \psi_\lambda^{\prime \prime}/\psi_\lambda ^\prime  + 2/r,	
\end{equation}
The dimensional constants $\lambda s^2$ cancel in $\psi_\lambda^{\prime \prime}/\psi_\lambda ^\prime  $; replacing $\psi_\lambda /\lambda s^2$ by the dimensionless $\Psi_\lambda $, one can write in (5.1),
\begin{equation}       \Psi_\lambda ^{\prime \prime}/\Psi_\lambda ^\prime  = d\log(|\Psi_\lambda ^\prime|)/dr			
\end{equation}
With the inclusion of the $2/r$ term from (5.1),
 \begin{equation}\mathcal{A} = d \log (s^2r^2 |\Psi _\lambda ^\prime |) /dr
\end{equation}
Using the dimensionless $\Phi_k \equiv \phi_ks^3/k$ in (4.11)
\begin{equation}d\Phi_k /dr + \Phi_k \mathcal{A} = Ks^3/r^2			
\end{equation}
Substituting for $\mathcal{A}$ in (5.4) from (5.3) and differentiating the log, (5.4) becomes,
\begin{equation}d \Phi_k/dr + \Phi_k [d(r^2|\Psi_\lambda^\prime |)/dr]/r^2\mid \Psi_\lambda^\prime | =  Ks^3/r^2	
\end{equation}
After multiplying across by $r^2 | \Psi_\lambda^\prime|$,
  \begin{equation}     
 d(\Phi_kr^2| \Psi_\lambda^\prime|)/dr 
 =  Ks^3| \Psi_\lambda^\prime|			
	\end{equation}
 		
\begin{equation} \Phi_k =  Ks^3\left[\int  | \Psi_\lambda^\prime|dr\right]/r^2| \Psi_\lambda^\prime|+ C/r^2| \Psi_\lambda^\prime|  			
\end{equation}
With 			   
\begin{equation}\left[ \int  | \Psi_\lambda^\prime|
dr\right]/r^2| \Psi_\lambda^\prime| =  
\left[\int \Psi_\lambda^\prime dr\right] /r^2 \Psi_\lambda^\prime		
\end{equation}
\begin{equation} \Phi_k= Ks^3(\Psi_\lambda + \beta)/r^2 \Psi_\lambda^\prime+ C/r^2| \Psi_\lambda^\prime| 	
\end{equation}
The constant $\beta$ (dimension $1/s^3$) consists of the integration constant plus the arbitrary constant added to $\Psi_\lambda $ (following (4.14)). Putting $Ks^3\beta = P$, one obtains the three distinct terms,	
\begin{equation} \Phi _k = Ks^3\Psi_\lambda/r^2\Psi_\lambda^\prime  + P/r^2\Psi_\lambda^\prime  +  C/r^2|\Psi_\lambda^\prime| 	
\end{equation}
Although $\Psi_\lambda$ is unknown we attribute to it the general property of having one or more extrema; with different $\Psi_\lambda$  and relative signs and magnitudes of the three dimensionless parameters $K, P$ and $C$, various multi-quark solutions follow from this general solution to the confining potential, including unstable particles [1]. All solutions of (5.10) diverge at the origin, and diverge asymptotically as $r^2$. The associated quarks form into a shell structure with one quark per shell--illustrated below.

For a 2-quark system,   $ \Psi_\lambda $ is needed with one maximum. As a simple example we consider a solution of (5.10) with $K = 0$ (making $P = 0$),
\begin{equation}
\Phi_k  = C/r^2|\Psi_\lambda^\prime |  				
\end{equation}

	The $\Phi_k$  divergence at the zero of $\Psi_\lambda^\prime$ divides space into two regions in each of which $\Phi_k$  is contained by two divergences, between the origin and the central divergence in the first region and between the central divergence and asymptotically in the second region. 
There is  one quark per region  near the minimum of the $\Phi_k$  there. The two corresponding $\phi_k$  potentials will in general have different values for the fractional charges and the confining charge $k$, which could be positive or negative. A simple $\Psi_\lambda$ with one maximum is illustrated below.
\begin{equation}\Psi_\lambda  = r/(s + \alpha r)^4
\end{equation}
\begin{equation}\Psi_\lambda  ^\prime  = (s - 3\alpha r)/(s + \alpha r)^5\end{equation}
\begin{equation}\Phi_k = (s + \alpha r)^5/r^2|s- 3\alpha r|			
\end{equation}
The dimensionless parameter $\alpha $ separates the scale factor $s$ from the quark position and is assumed to be very small, $\alpha  \ll 1$; the space divider divergence is at $r = s/3\alpha $. The mass of a quark consists of the minimum of the energy $\xi_p = b(\phi_k + \psi_\lambda )$ (4.9), plus the Coulomb energy. 
For the two quarks in a positively charged particle,
 \begin{equation}     \xi_p = b_1(\phi_{k1} + \psi_{\lambda 1}) + b_2(\phi_{k2} + \psi_{\lambda 2}), 	\end{equation}		 
where $b_1 = 2/3$ and $b_2 = 1/3$, or reverse, and $k_1, k_2, \lambda _1$ and $\lambda _2$ would all be positive to ensure a positive energy for each quark. For a neutral 2-quark system $b_1 = -b_2$; with $b_1$ negative say, $k_1$ and $\lambda _1$ would be negative.

	For a 3-quark system, a proton say, the weak potential $\psi_\lambda $ has one minimum and one maximum yielding two internal divergences that separate three 
quarks. One quark will have $b = -1/3$, the other two quarks with $b = 2/3$.

\section{Charge-Free Quarks}

	We hold to the view that ``charge is a manifestation of field'', held also by some physicists in the early years of the 20th century, including Max Born [2], but apparently not held universally. This is not to discount the importance of the traditional charge however. Without it a satisfactory theory of electromagnetism could not have been achieved.
  	
Previous attempts to modify Maxwell's equations where field replaced charge, were entirely ad hoc, and as expected, unsatisfactory. Now with a non-gauge invariant formulation there is the possibility of doing better. In a non-gauge invariant system gauge transformations provide an additional degree of freedom that can have physical relevance. The terms `gauge transformation' and `Coulomb potential' are well defined in Maxwell's equations, which are gauge invariant and have charge density as a source. Here the equations are neither gauge invariant nor will have a charge density. To avoid introducing new terminology at this time however the same terms will continue to be used.

	A  quote by Frank Wilczek [3] is pertinent. He regrets being urged to use the term Asymptotic Freedom instead of his preference, `Charge without Charge'. This quote is particularly apt here because the strong, as well as the weak, potentials are solutions of homogeneous equations where charge is not involved (4.9), (4.11). `Charge without Charge' can now also be shown to apply to electric charges in quarks using gauge transformations.
	
In (1.8), removing the external source and adding a gauge transformation with the scalar 
gauge function $Z$ one gets,	 
\begin{equation}
\partial_\mu [F_{\mu \nu} + q^{-1} (A_\mu  + \partial_\mu 
Z)(A_\nu  + \partial_\nu Z)]  =  0			
\end{equation}
Placing all terms involving the gauge function on the right hand side to replace the, unknown, external source,
\begin{equation}
\partial_\mu [F_{\mu \nu} + q^{-1} A_\mu A_\nu)  = 
 -q^{-1} \partial_\mu (A_\mu \partial _\nu Z + A_\nu \partial _\mu 
Z + \partial _\mu Z\partial _\nu Z)		
\end{equation}
The `source' now consists of the gauge terms 
interacting with the potential $A_\mu$ as well as separately, 
each with dimension of inverse length. Defining the dimensionless 
gauge field $\mathcal{Z} = q^{-1}Z$ to use in  (2.2), the equation is
 particularly simple,
\begin{equation}
{\rm  Div}[E - \phi(\mathcal{A} + d\mathcal{Z}/dr)] =  0 			
\end{equation}

\section{Gauge Transformation for Charged Particles}

A suitable choice of gauge transformation for 
(6.3) should lead asymptotically to a Coulomb potential 
representing a charge as well as a Coulomb potential for 
each of the quarks in a group of quarks. These requirements 
are met with the following gauge transformation,     
\begin{equation}
d\mathcal{Z}/dr = -3/2\mathcal{A}					
\end{equation}
Substituting (7.1) in (6.3),
\begin{equation} {\rm Div}[E - \phi (-1/2\mathcal{A})] =  0 	
\end{equation}
Thus, 			        	
\begin{equation}
 d\phi /dr -1/2\phi \mathcal{A} = 0  				
\end{equation}
Using (5.3) for $\mathcal{A}$ in (7.3) and dividing by $\phi$, the Coulomb 
charges cancel. In terms of the dimensionless Coulomb potential $\Theta$,
	
\begin{equation}
(d\Theta /dr)/\Theta   =  d\log(s\Theta )/dr  =  1/2d\log(s^2r^2
|\Psi_\lambda ^\prime |)/dr      		
\end{equation}
Thus, 
\begin{equation}			
\Theta = \pm  [\alpha r^2|\Psi_\lambda ^\prime |]^{1/2}
\end{equation}  				 
with integration constant $\alpha$. $\Theta$ has a zero for 
each extremum of $\Psi_\lambda  $ and is finite everywhere. 
Asymptotically $\Theta \sim  1/r$ and approaches the origin at least as fast as
 $r$.
The gauge function $\mathcal{Z}$ obtains easily by integrating (7.1) using (5.3),		
\begin{equation}	
\mathcal{Z} = \log(s^2r^2|\Psi_\lambda ^\prime |)^{-3/2} + \, \mbox{constant}
\end{equation}
Near the origin $\mathcal{Z}$ diverges at least as fast as $\log(s/r)^3$
 and asymptotically it diverges as $\log(r/s)^3$. Between there are 
the positive divergences from $|\Psi_\lambda ^\prime |$. The Coulomb potential (7.5) 
applies to any number of quarks depending on the number of extrema 
in $\Psi_\lambda $. Thus for a 3-quark particle, as discussed following (5.14),
 $\Psi_\lambda $ has both a minimum and a maximum producing two zeros in $\psi_\lambda ^\prime $
 that lead to two zeros in $\Theta$ and two divergences in $\Phi_k$.

\section{Coulomb Potential of a 2-Quark Particle}

In the first of the two regions, $\Theta$ is zero at the origin, remains positive and goes to zero at the end of the region where $|\Psi_\lambda ^\prime | = 0$.
 This part of the potential is completely enclosed inside the particle. It then starts at zero at the beginning of the second region, remains positive and decays asymptotically as $1/r$.  The charge-free potential functions essentially as a template 
for the dimensional fractional charges that can be accommodated 
now in each region of $\Theta$. Thus for a charged particle, the regions will share charges of $1/3$ and $2/3$, while for a neutral particle they will share either $1/3$ and $-1/3$ or $2/3$
 and $-2/3$.

	It is important to note that if the system were always static, only the Coulomb potential of the quark in the {\em outer} shell structure would be observed. But since one observes the total potential, time must move the fractional charges rapidly, at random, through the regions such that the time spent in each region is much shorter than the observation time--one sees then, effectively, both potentials 
`simultaneously'. The strong charge $k$, and the weak charge $ \lambda $, will move in concert with the Coulomb charges and maintain the integrity of the quarks.
Details for a 3-quark particle follow in a straightforward manner from those of the 2-quark particle.

	It is of interest to note that with the help of the gauge transformation one can refine the simple one maximum function $\Psi_\lambda $ 
(5.12) and its derivative (5.13). It reminds of a well-known quote, attributed to Einstein: ``everything should be made as simple as possible 
and not simpler.'' Thus for the Coulomb energy the 
spatial integral of $(d\phi /dr)^2$  should be finite for each quark, and the electric field from
 $d\phi /dr$ should be finite everywhere including the 
zero of $\Theta$. 
From (5.1), (7.3), (7.5), $d\phi /dr  = 
\frac{1}{2}\phi A \sim  |\Psi_\lambda ^\prime |^{1/2}\Psi_\lambda 
^{\prime \prime}  /\Psi_\lambda ^\prime $ near a zero of $ \Theta$. 
Hence, a simple one maximum $\Psi_\lambda $ with a zero of $\Psi_\lambda ^\prime $ that goes as $(s - \alpha r) ^n$, with arbitrary $n$, will have 
$\Psi_\lambda ^{\prime \prime}  /\Psi_\lambda ^\prime  \sim  
1/(s - \alpha r)$ and 
$d\phi /dr \sim  |(s - \alpha r) ^n|^{1/2}/(s - \alpha r)$.  
For $n  = 1$, as in (5.13), the electric field diverges and the Coulomb potential energy has a logarithmic singularity. So $n = 1$ is too simple. Less simple is using $n  = 3$ ($n$ must be odd for $\Psi_\lambda ^\prime $ to change sign 
crossing its zero), making $d\phi/dr$ and the Coulomb energy finite. 

	Not all mathematically acceptable gauge transformations may be of interest. Consider the gauge transformation,
\begin{equation}
d\mathcal{Y}/dr  =  -2\mathcal{A}\end{equation}
Applied to (6.3)), 
\begin{equation}
\nabla \cdot (E + \phi \mathcal{A})  =  0 
\end{equation}
The solution is now, (compare (7.5))
\begin{equation}
\Theta   = \pm [\alpha sr^2|\Psi_\lambda ^\prime |]  
\end{equation}
Asymptotically $\Theta \sim 1/r^2$ and goes as $r^2$ near the origin. This would represent a neutral quark, but an  
electrically neutral 
quark is not known to exist. The gauge transformation applied here (7.1) provides the desired replacement of electrical charge by field and so gives support to it as well as to the classical formulation, importantly within the context of a non-gauge invariant system.

\section{Conclusions}

	What stands out is the simple result for the Coulomb potential (7.5) in terms of the weak potential $ \Psi_\lambda$, which is also fundamental for the strong potential (5.10). In this classical formulation, $ \Psi_\lambda$ is the hub of the potentials in quarks---without a weak potential there would be no Coulomb or strong potential. 

The use of fields through gauge transformations in addressing the nature of electric charge in quarks has allowed for their complete and self contained description. The gauge transformation (7.1) may not be unique, but it is the simplest. 

	It is known that a particle acquires mass through the mechanism of the Higgs field; these results show that quarks in a hadron acquire electric charge through the agency of a gauge field via gauge transformations. That, and the appeal of simplicity, encourages the view that the gauge field may in fact turn out to be an expression of the Higgs field.

\section*{References}

\begin{enumerate}
\item [{[1]}] H.\ Schiff, {\em Int.\ Jour.\ of Theor.\ Phys: vol.\ 50}, Issue 2 (2011), p.\ 418. \\
arXiv:1012.1041v1 [math-phys] 5, Dec 2010.
\item  [{[2]}] Max Born, {\em Proc.\ Roy. Soc. A, vol. 143}, (1934), p.\ 410. 
\item [{[3]}] Frank Wilczek, {\em THE LIGHTNESS OF BEING}, Basic Books, (2008), p.\ 50. 
\end{enumerate}

\section*{Acknowledgement}
 
We would like to thank Werner Israel for discussions and friendly advice.

\end{document}